\documentclass[12pt]{article}
\usepackage{graphicx,cite,amssymb,epsfig,float,psfrag,multirow,amsmath,
subfigure}
\catcode`@=11

\def\@citex[#1]#2{\if@filesw\immediate\write\@auxout{\string\citation{#2}}\fi
  \def\@citea{}\@cite{\@for\@citeb:=#2\do
    {\@citea\def\@citea{,\penalty\@m}\@ifundefined
      {b@\@citeb}{{\bf ?}\@warning
       {Citation `\@citeb' on page \thepage \space undefined}}%
\hbox{\csname b@\@citeb\endcsname}}}{#1}}

\def\citer{\@ifnextchar [{\@tempswatrue\@citexr}{\@tempswafalse\@citexr[]}}

\def\@citexr[#1]#2{\if@filesw\immediate\write\@auxout{\string\citation{#2}}\fi
  \def\@citea{}\@cite{\@for\@citeb:=#2\do
    {\@citea\def\@citea{--\penalty\@m}\@ifundefined
       {b@\@citeb}{{\bf ?}\@warning
       {Citation `\@citeb' on page \thepage \space undefined}}%
\hbox{\csname b@\@citeb\endcsname}}}{#1}}
\catcode`@=12

\def\rt{{\tilde{r}}}
\def\zetat{{\tilde{\zeta}}}
\def\etab{{\bar{\eta}}}
\def\rob{{\bar{\rho}}}
\def\Spipi{{S_{\pi\pi}}}
\def\SKK{{S_{K K}}}
\def\CKK{{C_{K K}}}

\def\zt{\tilde{z}}
\begin{document}
 
\thispagestyle{empty}
\begin{flushright}
{\tt hep-ph/0407015}\\
{\tt LMU 05/04}\\
{\tt July 2004}
\end{flushright}

\vspace*{1.5cm}
{\centerline{\Large\bf Exploring  the Unitarity Triangle through} 
\centerline{\Large \bf $CP$ violation observables in  $B_s \to K^+ K^-$}}
\vspace*{2.cm}
\centerline{
{\sc A.~Salim~ Safir
}
}
\smallskip
\centerline{\sl Ludwig-Maximilians-Universit\"at M\"unchen, 
Sektion Physik,} 
\centerline{\sl Theresienstra\ss e 37, D-80333 Munich, Germany}

\vspace*{1.5cm}
\begin{abstract}
We discuss the determination of the CKM parameters from the forthcoming $CP$ violation observables in $B_s \to K^+ K^-$ decays. Combining the information on mixing induced $CP$ violation in $B_s \to K^+ K^-$, with the $B_d \to J/\psi K_s$ precision observable $\sin 2\beta$ and the $B^0_s$--$\bar{B^0_s}$ mixing phase $\phi_s$, we propose a determination of the unitarity triangle $(\bar\rho, \bar\eta)$. Computing the penguin parameters 
$(r, \theta)$ within QCD factorization yield precise determination of $(\bar\rho, \bar\eta)$, reflected by a weak dependence on the $\theta$ which is shown as a second order effect. The impact of the direct $CP$ violation observable $C_{KK}$ on the penguin parameters are investigated and a lower bound on $C_{KK}$ is extracted. We also discuss the effect of the 
$B^0_s$--$\bar{B^0_s}$ new physics mixing phase on the penguin parameters $(r, \theta)$ and $S_{KK}$. Using 
the $SU(3)$-flavour symmetry argument and the current $B$-factories data provided by the $B_d \to \pi^+ \pi^-$ modes, we complement the $B_s \to K^+ K^-$ $CP$-violating observables in a variety of ways, in particular we find that $S_{KK}>0$. Finally we analyze systematically the $SU(3)$-symmetry breaking factor within QCD factorization.
\vspace*{2.cm}

\end{abstract}

\newpage
\pagenumbering{arabic}


\section{Introduction}\label{Intro}
The two body non-leptonic transitions have played a very important role in 
exploring the $CP$ violation  through $B$-meson decays and in shaping both the sides and the three angles $\alpha$, $\beta$ and $\gamma$ of the Unitarity Triangle (UT) of the Cabibbo-Kobayashi-Maskawa (CKM) matrix \cite{CKM-C,CKM-KM} (for a detailed review, see \cite{Fleischer:2002ys}). In the
Wolfenstein parametrization \cite{Wolf} of the CKM matrix, characterized by 
the parameters $\lambda$, $A$, $\rho$ and $\eta$, the three inner angles of the UT are defined as:
\begin{eqnarray}\label{angles}
\sin (2\alpha)&=&\frac{2 \etab (\etab^2+\rob^2-\rob)}{(\rob^2+\etab^2)
((1-\rob)^2+\etab^2)},\nonumber\\
\sin (2\beta)&=&\frac{2 \etab (1-\rob)}{(1-\rob)^2+\etab^2},\qquad
\sin (2\gamma)=\frac{2 \etab \rob}{(\rob^2+\etab^2)},
\end{eqnarray}
where $\rob=\rho (1-\lambda^2/2)$ and $\etab=\eta (1-\lambda^2/2)$ are the perturbatively improved Wolfenstein parameters~\cite{BLO}.
Thanks to the precise measurements at the current $B$-factories, BABAR (SLAC) and Belle (KEK), $CP$ violation could recently be established with the help of the ``gold-plated'' mode $B_d \to J/\psi K_S$~\cite{Aubert:2001nu,Abe:2001xe}, 
leading to a precise measurement of $\sin2\beta$, where the current world average yields~\cite{HFAG} 
\begin{equation}\label{sin2bexp}
\sin 2\beta =0.739\pm 0.048.
\end{equation}

In contrast to the well measured CKM angle $\beta$, the other two angles 
$\alpha$ and $\gamma$ are poorly known at present and their determinations is 
more challenging~\cite{HQ}. Happily, the current thrust of the $B$-factories is now on the extractions of the other two angles, which will be measured through the $CP$ violation in the charmless $B$ decays, such as $B_d\to \pi\pi$ and similar modes~\cite{jawhery}. However, in this case
the extraction of weak phases is complicated by the so-called penguin pollution, leading to a hadronic model dependent estimate of the $CP$ asymmetries in these modes. An alternative approach to this problem is the use of symmetry arguments \citer{GL,Lavoura:2004rs}, such as the isospin or the $SU(3)$-symmetry, however this is very challenging from an experimental point of view.

In this paper we analyze the extraction of CKM parameters from the time-dependent $CP$ violation, both mixing-induced $(S)$ and direct $(C)$, in $B_s \to K^+ K^-$ decay, which is related to $B_d \to \pi^+ \pi^-$      by interchanging all down and strange quarks, i.e.~through the $U$-spin subgroup of the $SU(3)$-flavour-symmetry of strong interactions. To some extent theoretical input on the penguin-to-tree ratio will be needed and can be provided by the QCD factorization approach\cite{BBNS3}. However, we will show that the impact of uncertainties, especially those arising from the penguin phase, in the calculation is in fact negligeable. Moreover, even if the model-dependent assumptions on the penguin parameters are ignored, it is still possible to get useful informations on these parameters using $CP$ violation observables and the $B_s^0-\bar{B_s^0}$ mixing phase. Another possibility is the use of the current $B_d \to \pi^+ \pi^-$ experimental measurements via the $U$-spin symmetry in order to explore the future impact of the $B_s \to K^+ K^-$ on the extraction of CKM parameters. Although this strategy is not affected by any hadronic assumptions, its theoretical accuracy is  limited by $U$-spin-breaking corrections. 
In order to make uses of these methods, dedicated $B$ physics experiments at hadron colliders, Tevatron and LHC, will provide the adequate testing places for such decays. 

To make our analysis more transparent we propose the following strategy. After isolating systematically the hadronic quantities, namely the pure penguin 
amplitude $r$ and the strong phase $\theta$, from CKM parameters in the penguin-to-tree-ratio amplitude, we express the time-dependent $CP$ asymmetries in $B_s \to K^+ K^-$ in terms of Wolfenstein parameters $(\bar\rho, \bar\eta)$, hadronic quantities $(r, \theta)$ and the $B_s^0-\bar{B_s^0}$ mixing phase $\phi_s$. Combining them with the precise measurement of the ``gold-plated''      $B_d \to J/\psi K_S$ observable $\sin2\beta$, useful informations on the unitarity triangle, $(\bar\rho, \bar\eta)$, and penguin parameters could be extracted. To make our investigation more 
predictive, we use the current $B$-factories measurements on $B_d \to \pi^+ \pi^-$ modes to extrapolate the forthcoming physics potential of the $B_s\to K^+ K^-$ systems at future hadron machines. To illustrate this scenario, we have supplied from QCD Sum Rules the corresponding $SU(3)$-symmetry-breaking factor. To reinforce the validity of this approximation, we have tested systematically the $SU(3)$-breaking effects within the QCD factorization framework.
 
The utility of $B_s \to K^+ K^-$ to probe CKM phases was already the subject
     of extensive works~\citer{Dunietz:1993rm,Charles:2004jd}, where various strategies have been carried out in order to constrain theoretical uncertainties. Our new strategy proposed here consists of isolating the penguin quantities $r$ and $\theta$ from the CKM parameters in order to explore in a clean way the potential information of $CP$ violation observables, namely $\SKK$ and $\CKK$, combined with $\sin2\beta$ and the $B_s^0-\bar{B_s^0}$ mixing phase $\phi_s$ in the determination of the UT.

The outline of this paper is as follows. In Sec. 2, a completely general parameterization of the $CP$ violation observables in $B_s \to K^+ K^-$ is collected, 
in terms of pure hadronic quantities $(r, \theta)$ and the CKM parameters. Using the QCD factorization framework, we discuss the estimate of $B_s \to K^+ K^-$ penguin parameters in Sec.~3. In Sec.~4, we explore the determination of the UT, from both the mixing-induced and direct $CP$ violation parameters, $\sin2\beta$ and the $B_s^0-\bar{B_s^0}$ mixing phase $\phi_s$. The implications for the allowed contours in the $r$-$\theta$ planes fixed through the $CP$-violating observables are as well discussed. This alternative possibility is useful if new physics (NP)  affects the $B_s^0-\bar{B_s^0}$ mixing phase. Sec.~5 explores the implications of the current $B_d \to \pi^+ \pi^-$ $B$-factories measurements  
in extrapolating the physics potential of the $B_s\to K^+ K^-$ decays at future hadron colliders, using the $SU(3)$-flavour-symmetry.       This strategy is only affected by  the $SU(3)$-symmetry-breaking corrections. An estimate of these
 effects  is investigated in more detail within the QCD factorization framework. Finally, we summarize our results in Sec.~6.

\section{Basic Formulas}
The time-dependent $CP$ asymmetry in $B_s\to K^+ K^-$ decays
is defined by
\begin{eqnarray}\label{acppipi}
A^{K K}_{CP}(t) &=& 
\frac{B(B_s(t)\to K^+K^-)-B(\bar B_s(t)\to K^+ K^-)}{
  B(B_s(t)\to K^+ K^-)+B(\bar B_s(t)\to K^+ K^-)} \nonumber \\
&=& - S\, \sin(\Delta m_{B_s} t) + C\, \cos(\Delta m_{B_s} t),
\end{eqnarray}
where:
\begin{equation}\label{SandC}
S=\frac{2\, {\rm Im}\xi}{1+|\xi|^2},\qquad
C=\frac{1-|\xi|^2}{1+|\xi|^2},
\end{equation}
with
\begin{equation}\label{scxi}
\xi=e^{- i\,\phi_s}\,\frac{e^{-i\gamma}+P/T}{e^{+i\gamma}+P/T}
\end{equation}
defines the hadronic contribution in our decay mode. The phase $\phi_s\equiv -2 \delta\gamma=2\, \mathrm{arg}(V_{ts}^* V_{tb})$ denotes the $B_s^0-\bar{B_s^0}$ mixing phase. Within the Standard Model (SM), we have $2 \delta\gamma\simeq0.03$ due to a Cabibbo suppression of $\cal O$ $(\lambda^2)$, implying that $\phi_s$ is very small\footnote{Needless to note, a large measurement of $\delta\gamma$ much larger than the SM expectation of  $\cal O$$(0.03)$ would be a strong indication for NP contributions to $B_s^0-\bar{B_s^0}$ mixing.}, which could be extracted using the $B_s \to J/\psi \phi$ mode \cite{Dighe:1998vk}.
 
In terms of the Wolfenstein parameters $\bar\rho$ and $\bar\eta$
the CKM phase factors read
\begin{equation}\label{gambetre}
e^{\pm i\gamma}=\frac{\bar\rho\pm i \bar\eta}{\sqrt{\bar\rho^2+\bar\eta^2}}.
\end{equation}
On the other hand, the penguin-to-tree ratio $P/T$, defined in (\ref{scxi}), can be written as
\begin{equation}\label{ptrphi}
\frac{P}{T}=-\frac{r e^{i\theta}}{\epsilon\sqrt{\bar\rho^2+\bar\eta^2}}.
\end{equation}
where $\epsilon\equiv \lambda^2/(1-\lambda^2)$ and $\lambda=0.22$ is the Cabibbo angle.
The real parameters $r$ and $\theta$ defined in this way are
pure strong interaction quantities without further dependence
on CKM variables.

For any given values of $r$ and $\theta$ a measurement of $S$ 
defines a curve in the ($\bar\rho$, $\bar\eta$)-plane.
Using the relations above this constraint is given by the
equation
\begin{equation}\label{srhoeta}
S=
-\frac
{2\bar\eta (\bar\rho - \rt \cos\theta)\cos\phi_s +
(\rt^2+\bar\rho^2-\bar\eta^2 - 2 \rt \bar\rho \cos\theta)\sin\phi_s}
{\rt^2 + \bar\rho^2+\bar\eta^2 - 2 \rt\bar\rho \cos\theta},
\end{equation}
where $\rt\equiv r/\epsilon$ and similarly the relation between $C$ and 
($\bar\rho$, $\bar\eta$) reads
\begin{equation}\label{crhoeta}
C=-\frac{2 \rt \bar\eta\, \sin\theta}{
\rt^2 + \bar\rho^2+\bar\eta^2-2 \rt \bar\rho \cos\theta}.
\end{equation}

If there were no penguin amplitude $(\rt=0)$, then $\xi$ would be given by the pure phase $-\phi_s-2\gamma$. In that case, $C$ would vanish and $S$ would provide a clear determination of the phase $\gamma$. However, the presence of the penguin amplitude spoils this determination.
\section{Theoretical Framework}
In this section we discuss theoretical calculations
of the penguin contribution in $B_s\to K^+ K^-$.
At the quark level, the charmless hadronic decay $\bar b \to \bar s d \bar d$ can be described in terms of the effective weak Hamiltonian, obtained by integrating out the top quark and $W^{\pm}$ bosons:
\begin{equation}\label{heff}
{\cal H}_{eff}=\frac{G_F}{\sqrt{2}}\sum_{p=u,c}\lambda'_p
\left(C_1\, Q^p_1 + C_2\, Q^p_2 +\sum_{i=3,\ldots,10,\,7\gamma,\, 8g}C_i\, Q_i\right)
+ {\rm h.c.}
\end{equation}
where $Q_{1,2}^p$ are the left-handed current--current operators 
arising from $W^{\pm}$ bosons exchange, $Q_{3,\dots, 6}$ and $Q_{7,\dots, 10}$ are 
QCD and electroweak penguin operators, and $Q_{7\gamma}$ and $Q_{8g}$ 
are the electromagnetic and chromomagnetic dipole operators. They are 
given by

\begin{eqnarray}\label{operateurs}
   Q_1^p &=& (\bar p b)_{V-A} (\bar s p)_{V-A} \,,
    \hspace{2.5cm}
    Q^p_2 = (\bar p_i b_j)_{V-A} (\bar s_j p_i)_{V-A} \,, \\
   Q_3 &=& (\bar s b)_{V-A} \sum{}_{\!q}\,(\bar q q)_{V-A} \,,
    \hspace{1.7cm}
    Q_4 = (\bar s_i b_j)_{V-A} \sum{}_{\!q}\,(\bar q_j q_i)_{V-A} \,,
    \nonumber\\
   Q_5 &=& (\bar s b)_{V-A} \sum{}_{\!q}\,(\bar q q)_{V+A} \,, 
    \hspace{1.7cm}
    Q_6 = (\bar s_i b_j)_{V-A} \sum{}_{\!q}\,(\bar q_j q_i)_{V+A} \,,
    \nonumber\\
   Q_7 &=& (\bar s b)_{V-A} \sum{}_{\!q}\,{\textstyle\frac32} e_q 
    (\bar q q)_{V+A} \,, \hspace{1.11cm}
    Q_8 = (\bar s_i b_j)_{V-A} \sum{}_{\!q}\,{\textstyle\frac32} e_q
    (\bar q_j q_i)_{V+A} \,, \nonumber \\
   Q_9 &=& (\bar s b)_{V-A} \sum{}_{\!q}\,{\textstyle\frac32} e_q 
    (\bar q q)_{V-A} \,, \hspace{0.98cm}
    Q_{10} = (\bar s_i b_j)_{V-A} \sum{}_{\!q}\,{\textstyle\frac32} e_q
    (\bar q_j q_i)_{V-A} \,, \nonumber\\
   Q_{7\gamma} &=& \frac{-e}{8\pi^2}\,m_b\, 
    \bar s\sigma_{\mu\nu}(1+\gamma_5) F^{\mu\nu} b \,,
    \hspace{0.81cm}
   Q_{8g} = \frac{-g_s}{8\pi^2}\,m_b\, 
    \bar s\sigma_{\mu\nu}(1+\gamma_5) G^{\mu\nu} b \,,\nonumber
\end{eqnarray}
where $(\bar q_1 q_2)_{V\pm A}=\bar q_1\gamma_\mu(1\pm\gamma_5)q_2$, 
$i,j$ are colour indices, $e_q$ are the electric charges of the quarks 
in units of $|e|$, and a summation over $q=u,d,s,c,b$ is implied. The
definition of the dipole operators $Q_{7\gamma}$ and $Q_{8g}$ corresponds 
to the sign convention $iD^\mu=i\partial^\mu+g_s A_a^\mu t_a$ for the 
gauge-covariant derivative. The Wilson coefficients are calculated at 
a high scale $\mu\sim M_W$ and evolved down to a characteristic scale 
$\mu\sim m_b$ using next-to-leading order renormalization-group 
equations~\cite{BJLW}. The CKM factors are here denoted by 
$\lambda'_p=V^*_{ps}V_{pb}$.


\subsection{QCD Factorization}
When the QCD factorization approach~\cite{BBNS3} is applied to the charmless
decays $B_s\to K^+ K^-$, the hadronic matrix elements of the effective weak Hamiltonian can be written in the form:
\begin{eqnarray}\label{fact}
   \langle\ K^+\, K^-|{\cal H}_{eff}|\bar B_s\rangle \propto 
\frac{G_F}{\sqrt{2}} \sum_{p=u,c}\lambda'_p \bigg(a_i^p 
\langle\ K^+\, K^-|Q_i|\bar B_s\rangle_{nf} + f_{B_s} f_K^2\, b_i\bigg).
%
\end{eqnarray}
The above $\langle\ K^+\, K^-|Q_i|\bar B_s\rangle_{nf}$ are the factorized 
hadronic matrix elements, which have the same definitions as those in the 
naive factorization (NF) approach \cite{NF}. The ``nonfactorizable'' effects 
are included in the process dependent coefficients $a_i$ which include the 
${\cal O}(\alpha_s)$ corrections from hard gluon exchanges in the weak matrix 
elements, namely the vertex corrections and the              
hard spectator 
scattering. A new class of power corrections that need to be considered here 
are the uncalculable weak annihilation contributions $b_i$ in (\ref{fact}), 
which are parametrized as phenomenological quantities within the QCDF 
approach \cite{BBNS3}.  Explicit expressions for the QCD coefficients $a_i$ 
 and the  weak annihilation parameters $b_i$ can be found in \cite{BBNS3}.

Neglecting the very small effects from electroweak penguin contributions 
in (\ref{fact}), one can express the penguin parameter $r\, e^{i\theta}$ 
in the form \cite{BBNS3}
\begin{equation}\label{rqcd}
r\, e^{i\theta}= -
\frac{a^c_4 + r^K_\chi a^c_6 + r_A[b_3+2 b_4]}{
 a_1+a^u_4 + r^K_\chi a^u_6 + r_A[b_1+b_3+2 b_4]},
\end{equation}
where the quantities $r^K_\chi$ and $r_A$ are defined by
\begin{equation}\label{rchira}
r^K_\chi (\mu) = \frac{2 m^2_K}{\bar m_b(\mu)(\bar m_u(\mu)+\bar m_s(\mu))},
\qquad r_A=\frac{f_{B_s} f_K}{m^2_{B_s} F^{B\to K}_0(0)}.
\end{equation}
$r^K_\chi$ is defined in terms of the $\overline{MS}$ quark masses
$\bar m_q(\mu)$ and depends on the renormalization scale $\mu$.
$F^{B\to K}_0(0)$ is a $B\to K$ transition form factor, evaluated at momentum 
transfer $q^2=0$.

Both quantities in (\ref{rchira}) are formally of order $\Lambda_{QCD}/m_b$
in the heavy-quark limit. However, whereas $r_A\approx 0.003$ is indeed very 
small, $r^K_\chi(1.5\,{\rm GeV})\approx 0.8$ is numerically large. Consequently
weak annihilation effects are $\Lambda_{QCD}/m_b$ suppressed and  penguin 
contributions (including the uncalculable hard spectator terms)  
$a^{p=u, c}_6$ enhanced. Thus, an estimate of these ``chirally-enhanced'' corrections is necessary in exploring the phenomenology of nonleptonic $B$ decays.

\subsection{Numerical Analysis}\label{NA}
\renewcommand{\arraystretch}{1.3}
\begin{table}[t]
\begin{center}
\begin{tabular}{|c|c|c|c|c|c|c|c|c|}\hline\hline
&
$\mu$&
$m_s$&
$m_c$&
$f_{B_s}$&
$F_0^{B\to K}$&
$\alpha_{1,2}^{K}$&
$\lambda_{B_s}$&
$(\rho,\phi)_{H[A]}$
\\ \hline
  $r=$
& $\pm0.006$
& $\pm0.02$
& $\pm0.002$
& $\pm0.002$
& $\pm0.002$
& $\pm0.005$
& $\pm0.002$
&$\pm0.001$
\\ 
$0.11$
&
& 
& 
& 
& 
& 
& 
&$[\pm0.03]$
\\ \hline
 $\theta=$
& $\pm0.03$
& $\pm0.003$
& $\pm0.06$
& $\pm0.002$
& $\pm0.002$
& $\pm0.02$
& $\pm0.002$
&$\pm0.01$\\ 
$0.13$
& 
& 
& 
& 
& 
& 
& 
&$[\pm0.30]$
\\ \hline\hline

\end{tabular}
\vspace*{0.3cm}
\end{center}
\centerline{\parbox{14cm}{\it{
\caption{\label{tab:rphi}
Theoretical values for $r$ and $\theta$ and their
uncertainties from various sources within QCD factorization.
}}}}
\end{table}
As seen in Table \ref{tab:rphi} a theoretical estimate of the penguin 
parameters $r$ and $\theta$, from a calculation within the QCDF framework, 
is presented. 
We also investigate the sensitivity of our results on various sources of input
parameters, presented in Table \ref{tab:input}.
It turns out that two classes of uncertainties are found
\begin{itemize}
\item
The first one reflects, namely from the 2-8 columns, the uncertainties 
from input into the factorization formulas at next-to-leading order, 
as well as the sensitivity to the renormalization scale $\mu$. 
\item 
The second class of uncertainties is due to the model estimates employed 
for power corrections to hard spectator scattering ($H$) 
and weak annihilation effects ($A$), parameterized
by phenomenological quantities~\cite{BBNS3}
\begin{equation}\label{xhxa}
X_{H,A}=\left(1+\rho_{H,A}\,e^{i\phi_{H,A}}\right)\ln\frac{m_B}{\Lambda_h},
\end{equation}
where $\rho_{H,A}=0 \pm 1$, $\phi_{H,A}\in[0,\pi]$ and
the infra-red cut-off parameter $\Lambda_h=0.5\,{\rm GeV}$.
The impact on $r$ and $\theta$ of this second class of uncertainties
is seen to be
  completely dominated by the annihilation contributions, in the 
last column in Table \ref{tab:rphi}.
\end{itemize}

\renewcommand{\arraystretch}{1.3}
\begin{table}[t]
\begin{center}
\begin{tabular}{|c|c|c|c|c|c|}\hline\hline
$m_s(2 {\rm GeV})$&
$m_c(m_b)$&
$f_{B_s}$&
$F_0^{B\to K}(0)$&
$(\alpha_1,\alpha_2)^{K}$&
$\lambda_{B_s}$
\\ \hline
 $110\pm 25$
& $1.3\pm 0.2$
& $230\pm 30$
& $0.34\pm 0.05$
& $(0.2\pm 0.2,0.1\pm0.3)$
& $350\pm 150$
\\ \hline\hline
\end{tabular}
\end{center}
\centerline{\parbox{14cm}{\it{
\caption{\label{tab:input}
Input used for Table \ref{tab:rphi}. We take $\mu \in [m_b/2, 2 m_b]$.
The values for $m_c(m_b)$ are in GeV, whereas $m_s$,
$f_{B_s}$ and $\lambda_{B_s}$ are in MeV.  
}}}}
\end{table}
\noindent Adding the errors in quadrature we find
\begin{eqnarray}
r &=& 0.11 \pm 0.022 \pm 0.032,\\
\theta &=& 0.13 \pm 0.065 \pm 0.299,
\end{eqnarray}
where the first (second) errors are from the first
(second) class of uncertainties. Combining both in quadrature
we finally arrive at 
\begin{equation}\label{rphi}
r=0.11\pm 0.04, \qquad \theta=0.13\pm 0.31,
\end{equation}
which we take as our reference predictions for $r$ and $\theta$
in QCD factorization.

\section{
Exploring the UT with $B_s\to K^+ K^-$ modes}

\subsection{Determining $\bar{\rho}$ and $\bar{\eta}$}

In this section we discuss the determination of the unitariy
triangle by combining the information from $S$
with the value of $\sin 2\beta$, which is known with high 
precision from $CP$ violation measurements in $B\to J/\Psi K_S$ and 
the $B_s^0-\bar{B_s^0}$ mixing phase $\phi_s$.
As we shall see, this method allows for a particularly transparent
analysis of the various uncertainties.
Both $\bar\rho$ and $\bar\eta$ can be obtained, which fixes the
unitarity triangle. A comparison with other determinations then
provides us with a test of the Standard Model.

The angle $\beta$ of the unitarity triangle is given by
\begin{equation}\label{taus2b}
\tau\equiv\cot\beta=\frac{\sin 2 \beta}{1-\sqrt{1-\sin^2 2\beta}}.
\end{equation}
Using the current world average of $\sin 2\beta$ (\ref{sin2bexp}),
implies
\begin{equation}\label{tauexp}
\tau=2.26\pm 0.22.
\end{equation}
Given a value of $\tau$, $\bar\rho$ is related to $\bar\eta$
by
\begin{equation}\label{rhotaueta}
\bar\rho = 1-\tau\, \bar\eta.
\end{equation}
The parameter $\bar\rho$ may thus be eliminated from $S$
in (\ref{srhoeta}), which can be solved for $\bar\eta$ to yield
\begin{eqnarray}\label{etataus}
\bar\eta=\frac{(- 1 + \rt\,\cos\theta ) \,
     \Big( \cos\phi_s - \tau \,( S + \sin\phi_s ) \Big)-%
\sqrt{{\cal E}}}
{S + S\,{\tau }^2 - 2\,\tau \,\cos\phi_s + (-1 +\tau^2)\,\sin\phi_s},
\end{eqnarray}
where~:
\begin{eqnarray}\label{Efct}
{\cal E}&=&
{( 1 - \rt\,\cos\theta ) }^2\,
           {\Big( \cos\phi_s - 
               \tau \,( S + \sin\phi_s )  \Big) }^2\\
&&\hspace{-1.cm}-\Big[ ( 1 + \rt^2 - 
               2\,\rt\,\cos\theta ) \,( S + \sin\phi_s ) 
        ( S + S\,{\tau }^2 - 2\,\tau \,\cos\phi_s + 
               (-1 + {\tau }^2 )\,\sin\phi_s )  \Big]. \nonumber 
\end{eqnarray}

So far, no approximations have been made and the two expressions in 
(\ref{rhotaueta}) and (\ref{etataus}) are still completely general.
Once the theoretical penguin parameters $r$ and $\theta$ are provided, a 
straightforward determination of the CKM parameters $\bar\eta$ and $\bar\rho$
are obtained from the two observables $\tau$ (or $\sin 2\beta$) and $S$.
It is at this point that some theoretical
input is necessary. We will now consider the impact of the
parameters $r$ and $\theta$, and of their uncertainties, on the analysis.
\renewcommand{\arraystretch}{1.3}
\begin{table}[t]
\begin{center}
\begin{tabular}{|c|c|c|c|c|c|c|c|c|}\hline\hline
$S$&
$-0.8$&
$-0.6$&
$-0.4$&
$-0.2$&
$ 0.2$&
$ 0.4$&
$ 0.6$&
$ 0.8$
\\ 
\hline
 $\bar\eta$
& $-0.3$
& $-0.24$
& $-0.18$
& $-0.10$
& $0.16$
& $0.50$
& $1.17$
& $-4.89$
\\ \hline 
$\Delta\bar\eta$ 
& $-0.04$ 
& $-0.02$  
& $-0.01$  
& $-0.005$   
& $0.005$
& $0.007$
& $-0.0002$
& $0.012$
\\ \hline\hline
\end{tabular}
\end{center}
\centerline{\parbox{14cm}{\it{
\caption{\label{tab:etabaretcor}
The corresponding values of $\bar\eta$ and $\Delta\bar\eta$ respectively given in the expressions 
(\ref{etataus0}) and (\ref{deletatau}) as a function of $S$ for fixed $(r,\theta)=(0.11,0.13)$.
}}}}
\end{table}
We first would like to point out that the sensitivity of $\bar\eta$
in (\ref{etataus}) on the strong phase $\theta$  is rather mild. In fact,
the dependence on $\theta$ enters in (\ref{etataus}) only at second order.
Expanding in $\theta$ we obtain to lowest order\footnote{For simplicity, we have used $\phi_s= 0$, which is a good approximation in the SM, in deriving Eqs.~(\ref{etataus0}) and (\ref{deletatau}).}
\begin{equation}\label{etataus0}
\bar\eta=
-\frac{-1 + S\,\tau +{\sqrt{ 1 - S^2 }}}{\left( S - 2\,\tau  + S\,{\tau }^2 \right)}\,(\rt - 1).
\end{equation}
This result is corrected at second order in $\theta$ through
\begin{equation}\label{deletatau}
\Delta\bar\eta=\left(- 1 + S\,\tau  +
      \frac{-1 + S^2 + \rt\,{\left( 1 - S\,\tau  \right) }^2}
       {\left(- 1 + \rt \right) \,{\sqrt{1 - S^2}}} \right)
\frac{\rt\,{\theta}^2} {2\,
    \left( S - 2\,\tau  + S\,{\tau }^2 \right)},
\end{equation}
which is very attractive and consistent with the heavy-quark limit, where
the strong phase is formally either $\alpha_s$- or 
$\Lambda_{QCD}/m_b$-suppressed.
Due to the difficulty in estimating the strong phase, the expression in (\ref{deletatau}) permits a reasonable extraction of the CKM parameter $\bar\eta$, 
using a quantitative knowledge on the strong phase, as long it is of moderate 
size.
%
Since the penguin parameter $r$ is also small ${\cal O}(0.1)$, the second order effect from $\theta$ in (\ref{deletatau}) is even further reduced. To sketch its effect, we have illustrated in Table \ref{tab:etabaretcor} 
the impact of a variation of $S$ for $(r,\theta) = (0.11, 0.13)$ on the lowest order term $\bar\eta$ and its second order correction $\Delta\bar\eta$. 
\begin{figure}[t]
\psfrag{S}{$S$}
\psfrag{etab}{$\bar\eta$}
\begin{center}
\psfig{figure=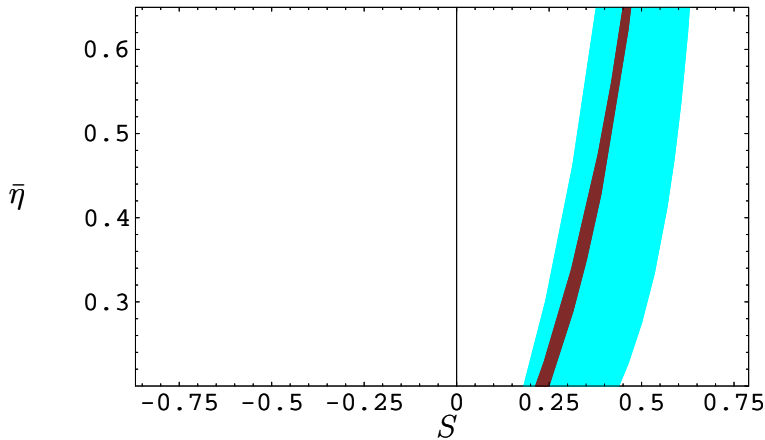}
\end{center}
{\it{\caption{CKM phase $\bar\eta$ as a function of the mixing-induced
$CP$ asymmetry $S$ in $B_s\to K^+K^-$ within the SM
for $\phi_s=-0.03$.
The dark (light) band reflects the theoretical uncertainty
in the penguin phase $\theta=0.13\pm 0.31$ 
(penguin amplitude $r=0.11\pm 0.04$).      
\label{fig:etabspp}}}}
\end{figure}

As an example, for $S=0.3$ and for typical values $r\approx 0.1$ this implies that 
$|\Delta\bar\eta|<0.015$ for $\theta$ up to $10^\circ$ and $|\Delta\bar\eta|<0.
11$ for relaxing the strong phase up to $30^\circ$, which is already a significant value. Consequently, the simple expression in (\ref{etataus0}) is most likely a good approximation to the exact one, requiring just minimal informations on the two $CP$ violating observables $S$  and $\tau$ in order to extract $\bar\eta$.

The determination of $\bar\eta$ as a function of $S$ is shown
in Fig. \ref{fig:etabspp},
which displays the theoretical uncertainty from the penguin
parameters $r$ and $\theta$ in QCD factorization. We note that QCD factorization 
prefers positive values of $S$. Furthermore, the sensitivity to the strong phase $\theta$ 
is less pronounced than for the penguin amplitude $r$, in extracting $\bar\eta$.

In the determination of $\bar\eta$ and $\bar\rho$ described
here discrete ambiguities do in principle arise.
One source is the well-known ambiguity in relating $\sin 2\beta$
to a value of $\beta$ (or $\tau=\cot\beta$).
Apart from the solution shown in (\ref{taus2b}), a second
solution exists with the sign of the square root reversed.
It corresponds to a larger value of $\beta$, incompatible
with the standard fit of the UT.
An additional ambiguity comes from the second solution
for $\bar\eta$, which is the result given in (\ref{etataus}) with a
positive sign in front of the square root.
This case may be considered separately, but will usually also yield
solutions in conflict with other information on the CKM phases.

\subsection{The Indirect $CP$  Violation}
In this section, we discuss the implications of the mixing-induced $CP$ violation, namely S, in the determination of the theoretical informations about the hadronic parameters $r$ and $\theta$.

\begin{figure}[t]
\psfrag{r}{$r$}
\psfrag{phi}{\hspace{-0.3 cm}$\theta$}
\psfrag{c}{\footnotesize{$(\phi_s=+\pi/9)$}}
\epsfig{file=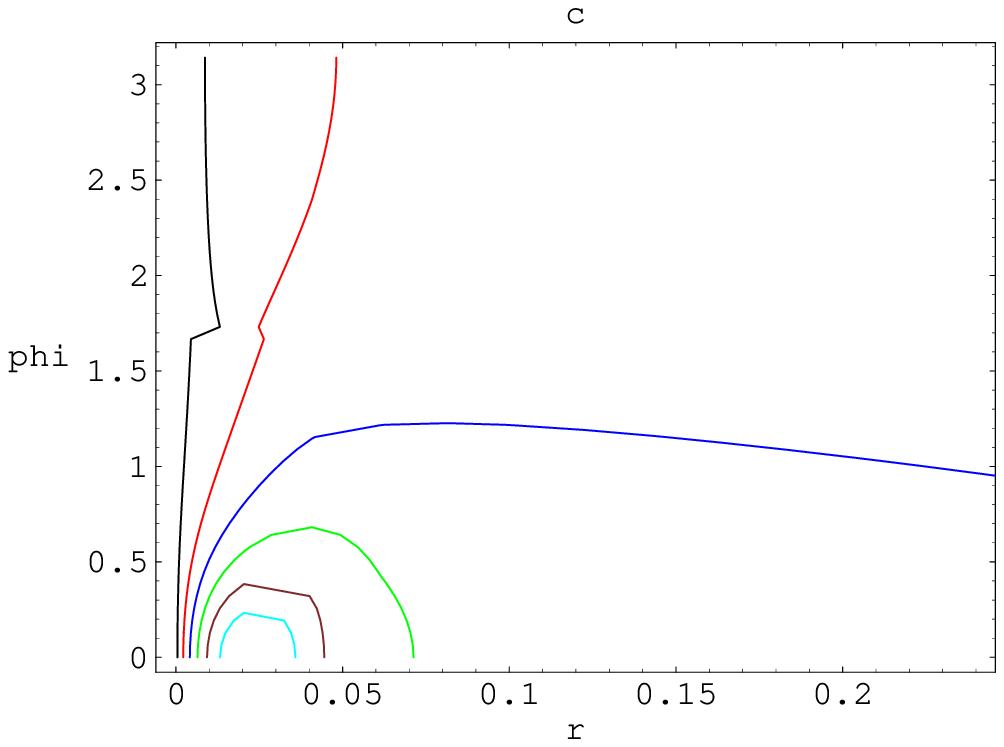,width=5cm,height=6cm}
\hspace{-0.5 cm}
\psfrag{phi}{}
\psfrag{b}{\vspace{-3 cm}\footnotesize{$(\phi_s=-0.03)$}}
\epsfig{file=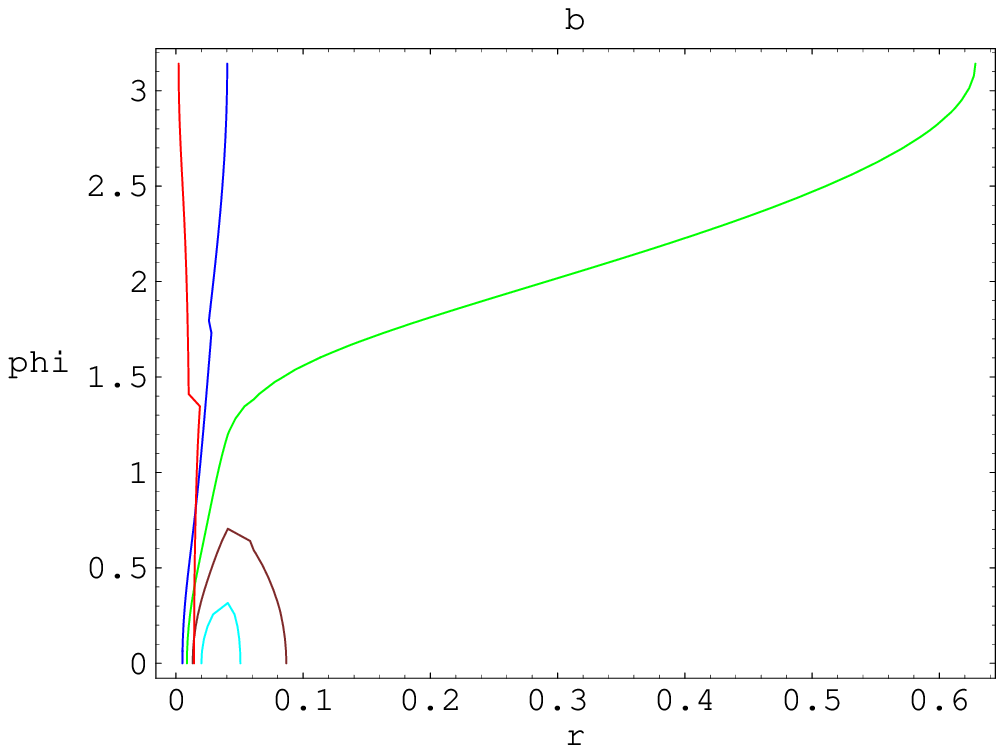,width=5cm,height=6cm}
\hspace{-0.5 cm}
\psfrag{phi}{}
\psfrag{e}{\footnotesize{$(\phi_s=-\pi/9)$}}
\epsfig{file=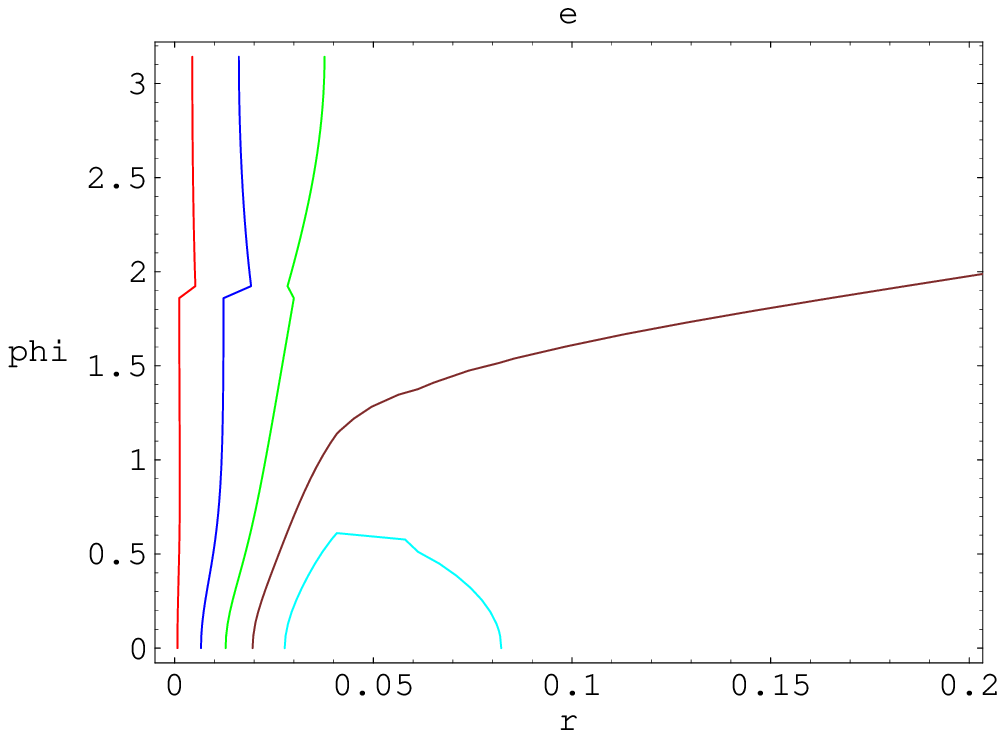,width=5cm,height=6cm}
\caption{\it 
Impact of a variation of $S$ within [-0.9,0,9] for $(\bar{\rho},\bar{\eta})=(0.20,0.35)$ 
and $\phi_s$ on the allowed ranges in the $(r-\theta)$ plane. 
These figures correspond from left to right, 
to $S$ = -0.9, -0.6, -0.3, 0, 0.3 and 0.6 for the case $\phi_s=+\pi/9$,
to $S$ = -0.6, -0.3, 0, 0.3 and 0.6 for both cases $\phi_s=-0.03$ and $\phi_s=-\pi/9$.}
\label{fig:phirS}
\end{figure}

Using (\ref{rhotaueta}), one can write (\ref{srhoeta}) in the form
\begin{eqnarray}\label{staueta}
S&=&\frac{ -1 + \etab\,\tau + \rt\,\cos\theta }
{\rt^2 + 1 - 2\,\etab\,\tau  + \etab^2\,( 1 + {\tau }^2) + 
    2\,\rt\,( -1 + \etab\,\tau  ) \,\cos\theta}
2\,\etab\,\cos\phi_s\nonumber\\
&&\hspace{.cm}-
\frac{
\rt^2 + 
   1 - 2\,\etab\,\tau  + 
     \etab^2\,\left( -1 + {\tau}^2 \right)  + 
       2\,\rt\,\left( -1 + \etab\,\tau  \right) \,\cos\theta}
{\rt^2 + 1 - 2\,\etab\,\tau  + \etab^2\,( 1 + {\tau }^2) + 
    2\,\rt\,( -1 + \etab\,\tau  ) \,\cos\theta}
\sin\phi_s.
\end{eqnarray}
In Fig. \ref{fig:phirS}, we illustrate the region in the $(r-\theta)$ plane that can be constrained  by the measurement of $S$ and the $B^0_s$--$\bar{B^0_s}$ mixing phase $\phi_s$. Within the SM this illustrates the correlation between the hadronic penguin parameters $(r,\theta)$ and the mixing induced $CP$ violation in $B_s \to K^+ K^-$ decays, as shown in the middle-plot of Fig. \ref{fig:phirS}. 
In the SM, the mixing phase $\phi_s$ is expected to be very small and hence any significant deviation from this correlation would be a striking signal of NP, entering through the modified  $B^0_s$--$\bar{B^0_s}$ mixing phase \cite{Grossman:1996er}. 

As an example, we have sketched in the left(right)-plot in Fig. \ref{fig:phirS} the scenario where $\phi_s=20^\circ~(-20^\circ)$, reflecting unambiguously the SM deviation. This means that, looking at Fig. \ref{fig:phirS}, it is possible to distinguish between the SM picture (the middle-plot) and some NP scenarios (the left- and right-one), which are 
contributing to the $B^0_s$--$\bar{B^0_s}$ mixing\footnote{Which is a 
loop-induced fourth order weak interaction process in the SM.} with some non-negligeable mixing phases. In this case, physics beyond the SM may also affect the $B_s \to J/\psi \phi$ decay amplitudes and compete with the SM tree-level amplitudes, however their relative impact is expected to be much smaller. On the other hand, NP signal in the  $B^0_s$--$\bar{B^0_s}$ mixing would be provided by a sizeable contributions either from the angular distributions analysis of $B_s\to J/\psi \phi$ and $B_s\to D^{*+}_s D^{*-}_s$ decays or from the $CP$-violating asymmetries 
in $B_s\to J/\psi \eta^{(')}$ and $B_s\to D^{+}_s D^{-}_s$ systems. An alternative discussion on this issue can be found in \cite{Dunietz:2000cr,DL}.

Since the $\phi_s$ is very small in the SM, $S$ is well approximated by:
\begin{eqnarray}\label{srbgam}
S=
\frac{2\,\etab\,\left( -1 + \etab\,\tau  +
      \rt\,\cos\theta \right) }{1 + \rt^2 + 
    \etab\,\left( \etab - 2\,\tau  + \etab\,{\tau}^2 \right)  + 
    2\,\rt\,\left( -1 + \etab\,\tau  \right) \,\cos\theta}.
\end{eqnarray}
Taking \cite{CKMF}
\begin{equation}\label{tauetaexp}
\tau=2.26\pm 0.22,\qquad \bar\eta=0.35\pm 0.04,
\end{equation}
and our penguin parameters results in (\ref{rphi}), we find from (\ref{staueta}) that 
\begin{equation}\label{spred}
S= +0.35 \quad ^{-0.01}_{+0.01}\,(\tau)
         \quad ^{+0.02}_{-0.02}\,(\bar\eta)
         \quad ^{-0.08}_{+0.18}\, (r)
         \quad ^{-0.04}_{-0.00}\,(\theta).
\end{equation}
We note that the uncertainty from the $\tau$ or $\sin 2\beta$, which reflects 
the current experimental accuracy in this quantity, is negligible ($\sim 3\%$) and the uncertainty from $\bar\eta$ , which for the purpose of predicting $S$ 
has here been taken from a standard CKM fit, is rather moderate ($\sim 7\%$).

The error from the hadronic phase $\theta$ is considerably larger ($\sim 10\%$).
 However, the dominant uncertainty, is due to the penguin parameter $r$ (up to $\sim 50\%$). 
The theoretical prediction in (\ref{spred}) shows that the SM prefers positive values for $S$, but it is difficult to obtain an accurate prediction.

\subsection{The Direct $CP$ Violation}
\begin{figure}[t]
\psfrag{rKK}{$r$}
\psfrag{phiKK}{$\theta[rad]$}
\begin{center}
\psfig{file=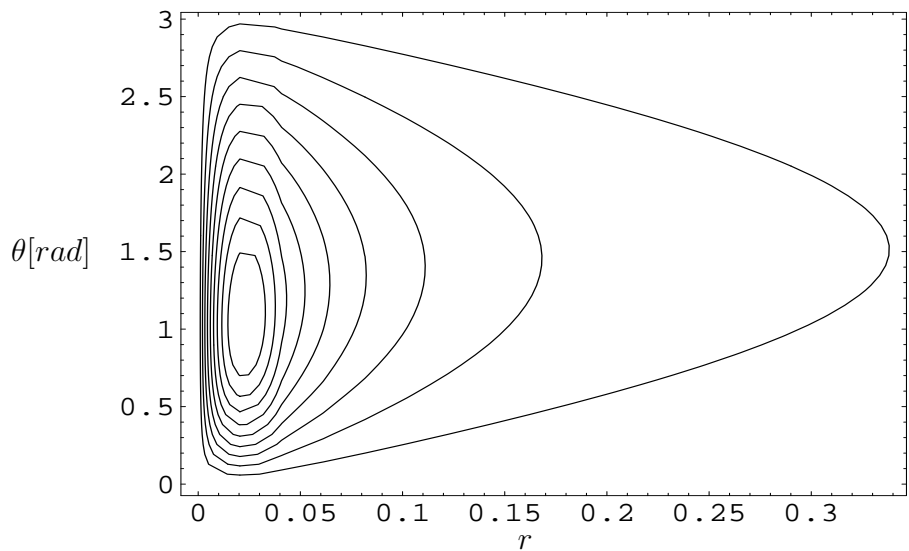,width=12cm,height=08cm}
\end{center}
{\it{\caption{
Contour of constant C in the $(r,\theta)$-plane for fixed $(\bar{\rho},\bar{\eta})=(0.20,0.35)$. 
These contours correspond from right to left, to $C$=-0.1,-0.2,-0.3,-0.4,-0.5, -0.6,-0.7,-0.8 and -0.9.  
\label{fig:phir}}}}
\end{figure}
Whereas so far in this section, we have focused on the implications of the mixing-induced $CP$ violation observable $S$, we now investigate how useful additional information can be extracted from a measurement of the direct $CP$ violation
parameter $C$.

The observable $C$ (given in Eq.~(\ref{crhoeta})) is independent 
on the $B_s^0-\bar{B_s^0}$ mixing phase $\phi_s$, contrary to 
the indirect $CP$  violation observable $S$, reflecting an insensitivity to 
NP contributions which could affect the $B_s^0-\bar{B_s^0}$ mixing. 
Since $C$ is an odd function of $\theta$, it is therefore sufficient 
to restrict the discussion to positive values of $\theta$, as shown in 
Fig.~\ref{fig:phir}. A positive 
phase $\theta$ is obtained by the perturbative
estimate in QCD factorization, neglecting soft phases with power
suppression. For positive $\theta$ the observable $C$ will be negative,
assuming $\bar\eta > 0$, and a sign change in $\theta$ will simply
flip the sign of $C$. Needless to note, in Fig.~\ref{fig:phir} no 
approximation has been made\footnote{the only input parameters are 
$(\bar{\rho},\bar{\eta})=(0.20,0.35)$, fixed through the actual CKM 
Fit~\cite{CKMF}.}, which shows a model-independent correlation 
between  the penguin parameters, $r$ and $\theta$, for different values for 
$C$ and any deviations from these predictions should be understood more as 
a strong indication of an uncontrolled non-perturbative contributions than a 
NP signal. 

In contrast to $B_d \to \pi^+ \pi^-$ modes, 
the hadronic quantities $r$ and $\theta$ are less pronounced for the direct $CP$ violation in $B_s\to K^+ K^-$ than for the former ones~\cite{BS-preparation}. In the SM for fixed $(\bar{\rho},\bar{\eta})=(0.20,0.35)$, a model-independent correlation between $C$ and the hadronic parameters $(r,~\theta)$ implies $r_{max}\approx 0.34$ (or $\rt_{max}\approx 7$) for $C= -0.1$ and $\theta=\pi/2$. Since $C$ is an odd function of $\theta$, $C=0.1$ and $\theta=-\pi/2$ implies as well $r_{max}$. In $B_d \to \pi^+ \pi^-$ decays, this would imply\footnote{The model-independent correlation between $C_{\pi\pi}$ and the hadronic parameters 
$(r,~\theta)_{\pi\pi}=(r',~\theta')$ would even allow the case $r'> 1$. } $r'_{max} = 1$ for $C_{\pi\pi}<0.8$ independently of $\theta$~\cite{BS-preparation}. 

As it has been shown in \cite{BS-preparation} for $B_d\to \pi^+ \pi^-$ cases, 
a bound on the $B_s\to K^+ K^-$ direct $CP$ violation parameter $C$ as well exists. This is straightforwardly obtained by minimizing $C$ with respect to the weak phase $\gamma$. Hence, we get:
\begin{equation}\label{barc}
C_{max}=\frac{-2 \zt\, \sin\theta}{
  \sqrt{(1+\zt^2)^2 -4 \zt^2 \, \cos^2\theta}},
\end{equation}
where the maximum occurs at $\cos\gamma= 2\zt\cos\theta/(1+\zt^2)$, with\begin{equation}\label{kappat}
\zt \equiv \frac{\rt}{R_b}=\bigg|\frac{P}{T}\bigg|,\,\,\,\,\,\,\,\,\,\,\,\,\,\,\,
R_b \equiv \sqrt{\bar\rho^2 + \bar\eta^2}.
\end{equation}
Contrary to $B_d\to \pi^+ \pi^-$ modes where $z (\equiv|P/T|_{\pi\pi})\leq 1$, the $B_s\to K^+ K^-$ decay prefers the $\zt \geq 1$ scenario. 
Intuitively this could be understood by the fact that $\zt$ (or $|P/T|$) is a doubly Cabibbo-enhanced term. Then, If $\zt=1$, or equivalently $\rt=R_b$, then $C_{max}\equiv -1$ independent
of $\theta$, and no useful upper bound is obtained.
On the other hand, if $\zt > 1$, then $C_{max}$ is maximized for
$\theta=\pi/2$. Under the general assumptions stated above and
without any assumption on the strong phase $\theta$ we thus find
the general bound
\begin{equation}\label{cbound0}
C >- \frac{2\zt}{1+\zt^2}.
\end{equation}
For the conservative bound $r \approx 0.15$
, $\zt \approx 7.69$ this implies
$C \gtrsim - 0.26$, which is already a strong constraint on this parameter. The bound on $C$ can be strengthened by using information
on $\theta$, as well as on $\zt$, and employing (\ref{barc}).
Then $\zt \approx  7.69$ and $\theta < 45^\circ~(30^\circ)$ gives 
$C \gtrsim -0.18~ (-0.13)$.

\section{The $SU(3)$-flavour-symmetry breaking in $B_s\to K^+ K^-$ 
vs. $B_d\to\pi^+ \pi^-$ decay}

The analyses described above concern the impact of the $CP$-violating observables in the $B_s\to K^+ K^-$ system, combined with the precise measurement of $\sin2\beta$,  in the extraction of the CKM parameters $\bar\rho$ and $\bar\eta$.

However a precise estimate of the CKM parameters, require theoretical input on the penguin parameter $(r,\theta)$. The first possibility could be based on the QCD factorization estimate of this parameters assuming the control over the non-perturbative parameters, namely the subleading effects. The second possibility is to use actual experimental informations on the $B_d\to \pi^+ \pi^-$ system in order to estimate our penguin parameter $(r,\theta)$, using the $SU(3)$-flavour-symmetry argument, on which we will focus in this section.
\subsection{Determining $(\bar\rho,\bar\eta)$ from $B_s\to K^+ K^-$ and $B_d\to \pi^+ \pi^-$}
Since the $B_s\to K^+ K^-$ decay could be related to the $B_d\to \pi^+ \pi^-$ one, by interchanging at the quark-level the $s$-quark by the $d$-quark, one can easily write in a similar way (\ref{srhoeta}) and (\ref{crhoeta}) for the $B_d\to \pi^+ \pi^-$ channel.

Using a notation similar to that in (\ref{ptrphi}), we obtain in terms of the corresponding penguin parameters $(r',\theta')$\cite{BS-preparation,S-HEP}:
\begin{equation}\label{srhoetapipi}
\Spipi=\frac{2\bar\eta [\bar\rho^2+\bar\eta^2-r'^2-\bar\rho(1-r'^2)+
       (\bar\rho^2 +\bar\eta^2-1)r' \cos\theta']}{((1-\bar\rho)^2+\bar\eta^2)
         (\bar\rho^2+\bar\eta^2+r'^2 +2 r'\bar\rho \cos\theta')}.
\end{equation}
Similarly the relation between $C_{\pi\pi}$ and ($\bar\rho$, $\bar\eta$) reads:
\begin{equation}\label{crhoetapipi}
C_{\pi \pi}=\frac{2 r'\bar\eta\, \sin\theta'}{
   \bar\rho^2+\bar\eta^2+r'^2 +2 r'\bar\rho \cos\theta'}.
\end{equation}
The parameter $\bar\rho$ may thus be eliminated from $S_{\pi \pi}$
in (\ref{srhoetapipi}), which can be solved for $\bar\eta$ to yield at the lowest order\footnote{As it has been noticed in \cite{BS-preparation,S-HEP}, the sensitivity of $\bar\eta$ on the strong phase $\theta'$ is rather mild, and it  enters  only at the second order.} in $\theta'$ \cite{BS-preparation,S-HEP}
\begin{equation}\label{etataus0pipi}
\bar\eta\dot=\frac{1+\tau \Spipi -\sqrt{1-\Spipi^2}}{(1+\tau^2)\Spipi}(1+r'),
\end{equation}
which is the analogue to (\ref{etataus0}) for the $B_s\to K^+ K^-$ system. Since the decays $B_d\to \pi^+ \pi^-$ and $B_s\to K^+ K^-$ are related to each other by interchanging all strange and down quarks, the $SU(3)$-flavour-symmetry of strong interactions implies:
\begin{eqnarray}
r&=&r',\label{rSU3}\\
\theta&=&\theta'.\label{thetaSU3}
\end{eqnarray}

Assuming that the $B_s^0-\bar{B_s^0}$ mixing phase $\phi_s$ is negligibly small, the weak strong phase dependence in $\bar\eta$ and considering the 
$SU(3)$-symmetry-breaking effects, one obtains from (\ref{etataus0}) and (\ref{etataus0pipi}):
\begin{eqnarray}\label{rSU3}
r'=
\frac{\frac{1 +\tau \Spipi- \sqrt{1 - \Spipi^2}} 
{\Spipi\, \left(1 + {\tau}^2 \right)} +
      \frac{1 - \tau \SKK - \sqrt{1 - \SKK^2}} 
{\SKK - 2\, \tau + \SKK\, \tau^2}}
{
      \zetat_{SU3} \frac{ 1 - \tau \SKK - \sqrt{1 - \SKK^2}}
      { \SKK - 2\, \tau + \SKK\, {\tau}^2 }
-
\frac{1 +\tau \Spipi- \sqrt{1 - \Spipi^2}}{
              \Spipi\, \left(1 + {\tau}^2 \right)}
},
\end{eqnarray}
where $\zetat_{SU3}=\zeta_{SU3}/\epsilon$ and:
\begin{eqnarray}\label{fa-SU3}
\zeta_{SU3}\equiv\frac{r}{r'}=1\pm 0.3, 
\end{eqnarray}
represents our $SU(3)$-symmetry-breaking estimate in relating the hadronic physics of our corresponding modes, namely $B_s\to K^+ K^-$ and $B_d\to \pi^+ \pi^-$. The 
reasonable value quoted in (\ref{fa-SU3}) is chosen as an {\it educated guess} for our analyses.
\begin{figure}[t]
\hspace{-2.cm}
\psfrag{r}{$r'$}
\psfrag{SKK}{$S_{K K}$}
\begin{center}
\epsfig{file=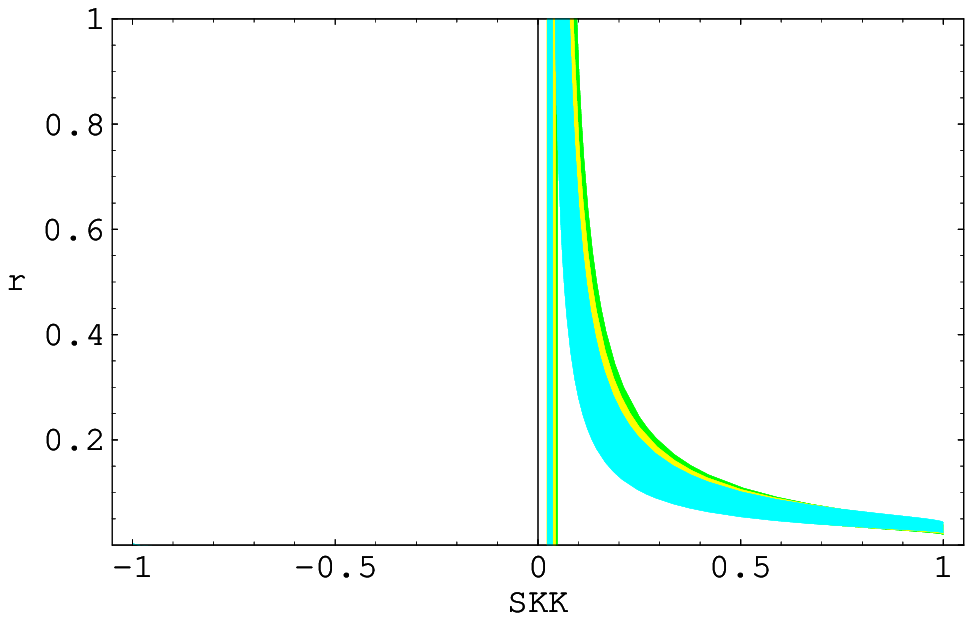,width=7.5cm,height=5cm}
\hspace{-0.75cm}
\psfrag{r}{}
\epsfig{file=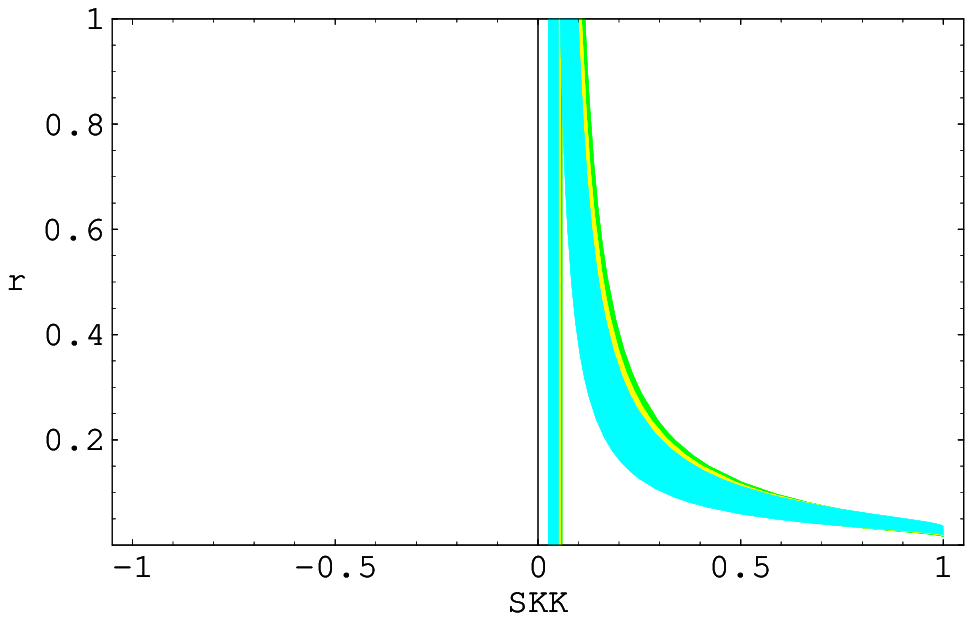,width=7.5cm,height=5cm}
\end{center}
{\it{\caption{
The dependence of $r'$ on $S_{K K}$ fixed through  the $CP$-violating 
observables $\Spipi$ and the $SU(3)$-symmetry-breaking factor 
$\zeta_{SU(3)}=1\pm0.3$. The band reflects the theoretical estimate on the  
$SU(3)$-symmetry-breaking
for the corresponding $\Spipi$. 
In the left figure, the bands correspond, from bottom to top, to 
$\Spipi=-0.8, -0.5 $ and $-0.1$. However in the right one, 
from bottom to top, to $\Spipi= 0.1,~0.5$ and $0.8$.
\label{fig:rSKK}}}}
\end{figure}

In Fig.~\ref{fig:rSKK}, we have plotted the dependence of $r$ on $S_{K K}$, for various mixing-induced $CP$ violation observable $S_{\pi \pi}$, using our 
$SU(3)$-symmetry-breaking estimate defined in (\ref{fa-SU3}). We observed that constraining our penguin parameter $r'$ (and hence $r$) to be positive implies a positive value of $S_{K K}$, in agreement with the result obtained recently 
by the CKMfitter Group in \cite{Charles:2004jd}.

Above, we have presented the dependence of $r$ on $S_{K K}$, for various mixing-induced $CP$ violation observable $S_{\pi \pi}$, using our $SU(3)$-symmetry-breaking effects, which could be helpful in estimating the penguin parameter $r$ once the experimental measurements on $B_s\to K^+ K^-$ will become available. In the following we shall investigate how useful information could we get from $S_{\pi \pi}$ and $\bar\eta$ in predicting $S_{K K}$. For this task, it is convenient, using (\ref{etataus0}) and (\ref{etataus0pipi}), to express $\bar\eta$ in terms of $S_{K K}$, $S_{\pi \pi}$ and $\tau$, getting :
\begin{figure}[t]
\hspace{-2.cm}
\psfrag{etab}{\hspace{0.2cm}$\bar\eta$}
\psfrag{SKK}{$S_{K K}$}
\begin{center}
\epsfig{file=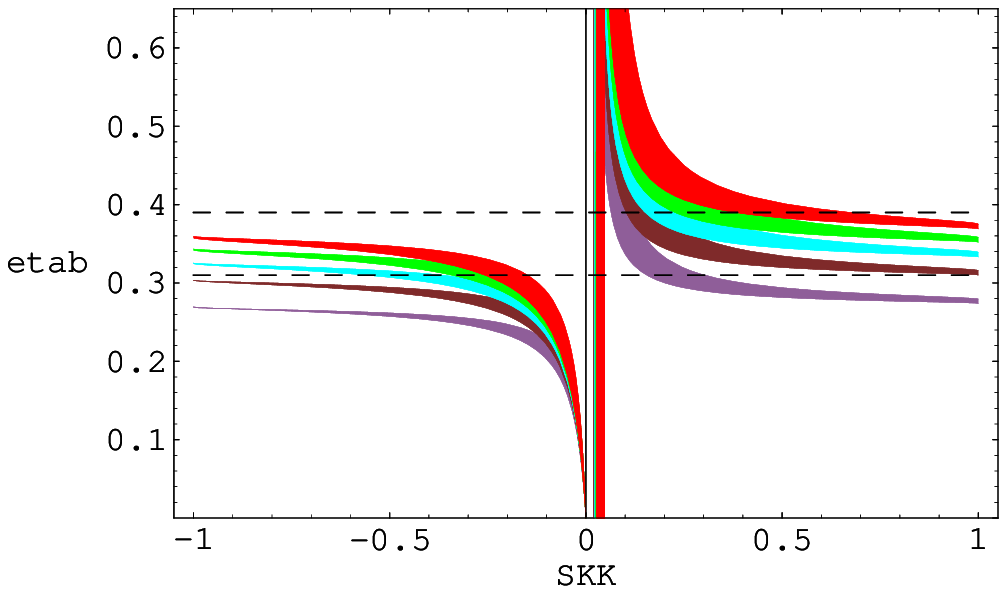,width=7.5cm,height=5cm}
\hspace{-0.75cm}
\psfrag{etab}{}
\epsfig{file=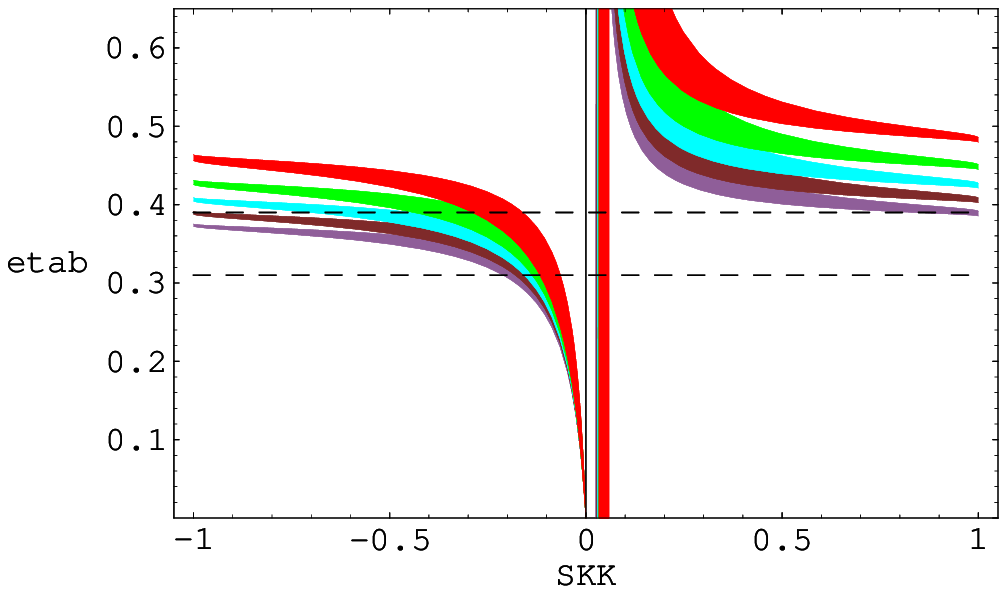,width=7.5cm,height=5cm}
\end{center}
{\it{\caption{
The dependence of $\bar\eta$ on $S_{K K}$ fixed through  the $CP$-violating observables $S_{\pi \pi}$ and the $SU(3)$-symmetry-breaking factor $\zeta_{SU(3)}=1\pm0.3$. 
The bands reflect the theoretical estimate on the  $SU(3)$-symmetry-breaking.
In the left figure, the bands correspond, from bottom to top, to 
$\Spipi=-0.9, -0.7, -0.5, -0.3$ and $-0.1$. However in the right one, from bottom to top, to $\Spipi=
0.1,~0.3,~0.5,~0.7$ and $0.9$.
The dashed lines delimit the actual $\pm1\sigma$ CKM Fit range on the $\bar\eta$ \cite{CKMF}.
\label{fig:etaSKK}}}}
\end{figure}

\begin{eqnarray}\label{etabSU3}
\bar\eta=
\frac{\zetat_{SU3} + 1}
{\Bigg(
\frac{\zetat_{SU3}\,( 1 + \tau^2)\,\Spipi}{1+\tau \Spipi-\sqrt{1 - \Spipi^2}}-
\frac{- 2\,\tau + ( 1 + \tau^2)\,\SKK }{1 - \tau \SKK - \sqrt{1 - \SKK^2}}
\Bigg)}.
\end{eqnarray}

Eq.~(\ref{etabSU3}) is very attractive, since it allows to extract the 
incoming information concerning the $B_s\to K^+ K^-$ system at the future
hadron 
machines, using the actual $B$-factories data on the $B_d\to \pi^+ \pi^-$ mode. For that one has to make an important assumption based on the $SU(3)$-flavour-symmetry, to relate the corresponding penguin parameters in the two 
systems.
\begin{figure}[t]
\psfrag{Coeff}{}
\psfrag{SKK}{}
\psfrag{Spipi}{{\scriptsize$\Spipi=-0.7$}}
\psfrag{C0}{\vspace{-0.7cm}$\kappa_o$}
\psfrag{C1}{\vspace{0.5cm}$\kappa_{1}$}
\psfrag{C2}{\vspace{0.7cm} $\kappa_2$}
\begin{center}
\epsfig{file=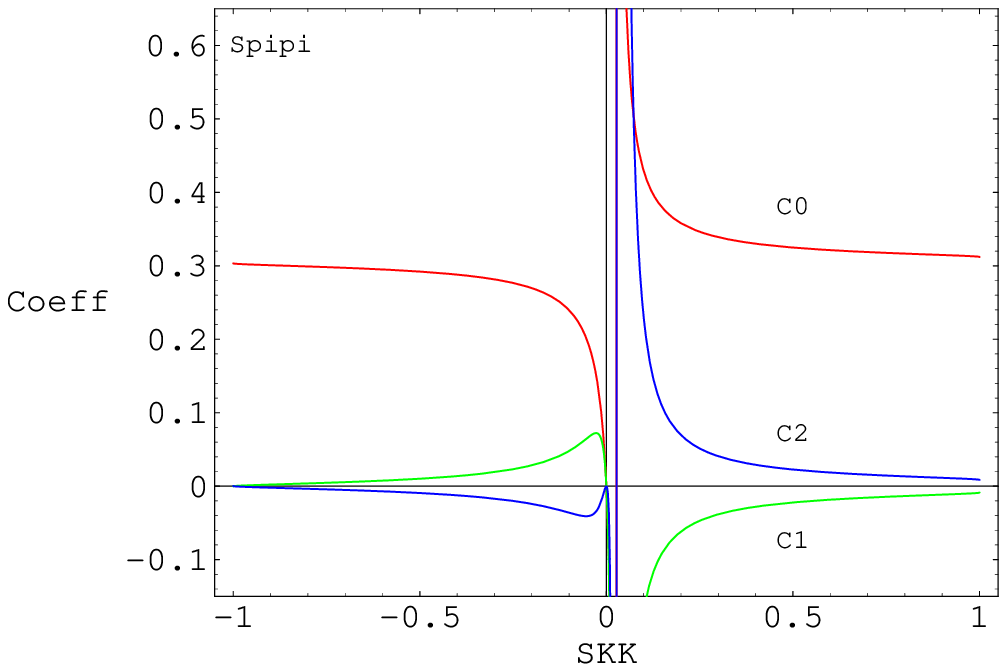,width=7.5cm,height=5cm}
\hspace{-1.cm}
\psfrag{Coeff}{}
\psfrag{Spipi}{\scriptsize{$\Spipi=-0.3$}}
\epsfig{file=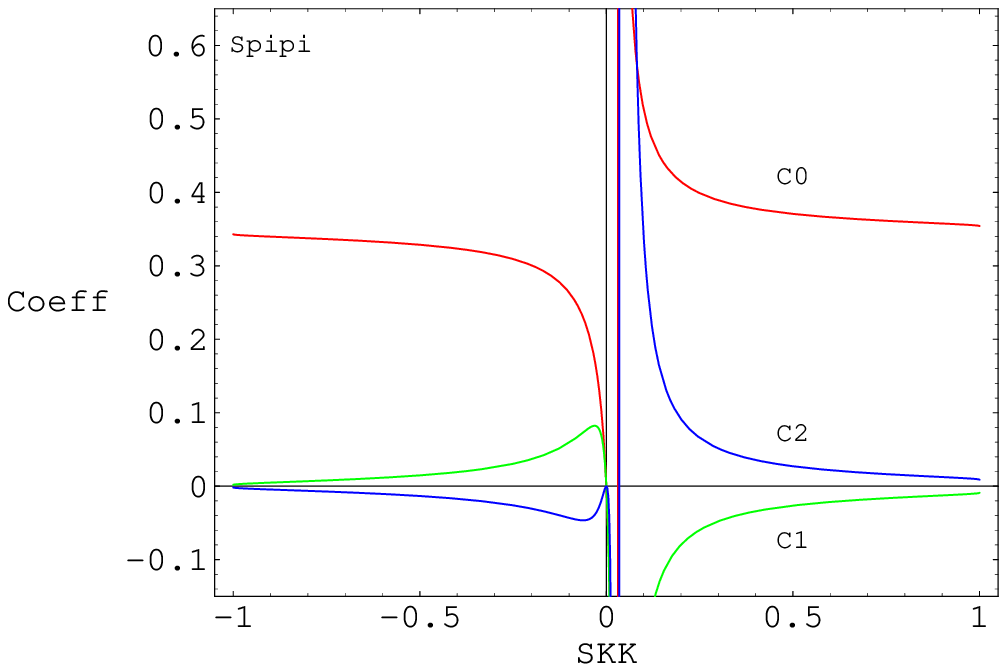,width=7.5cm,height=5cm}
\psfrag{SKK}{$S_{K K}$}
\psfrag{Spipi}{\scriptsize{$\Spipi=0.7$}}
\epsfig{file=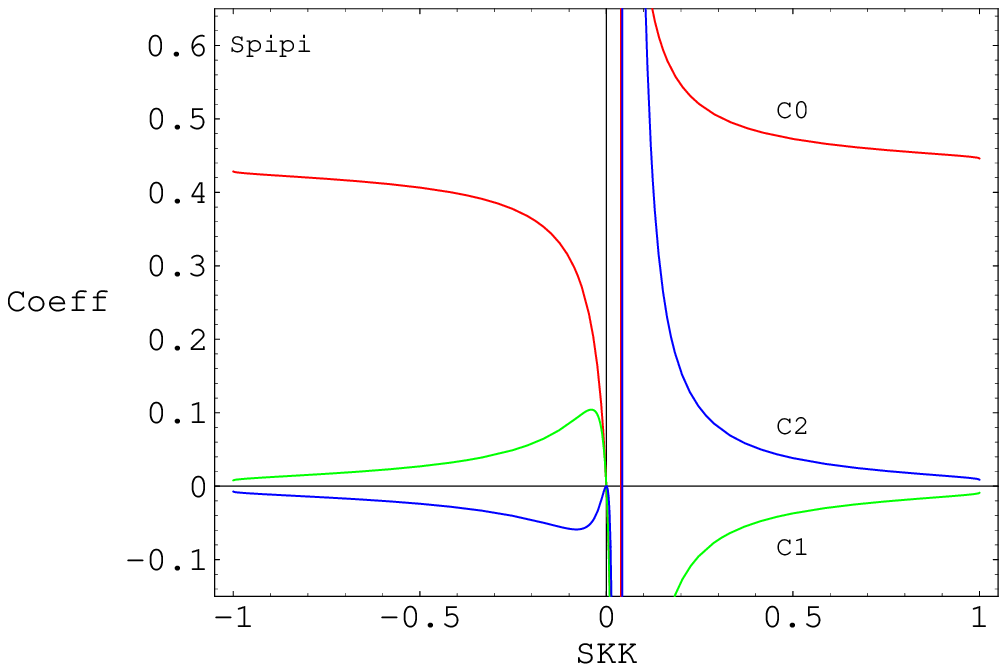,width=7.5cm,height=5cm}
\hspace{-1.cm}
\psfrag{Spipi}{\scriptsize{$\Spipi=0.3$}}
\epsfig{file=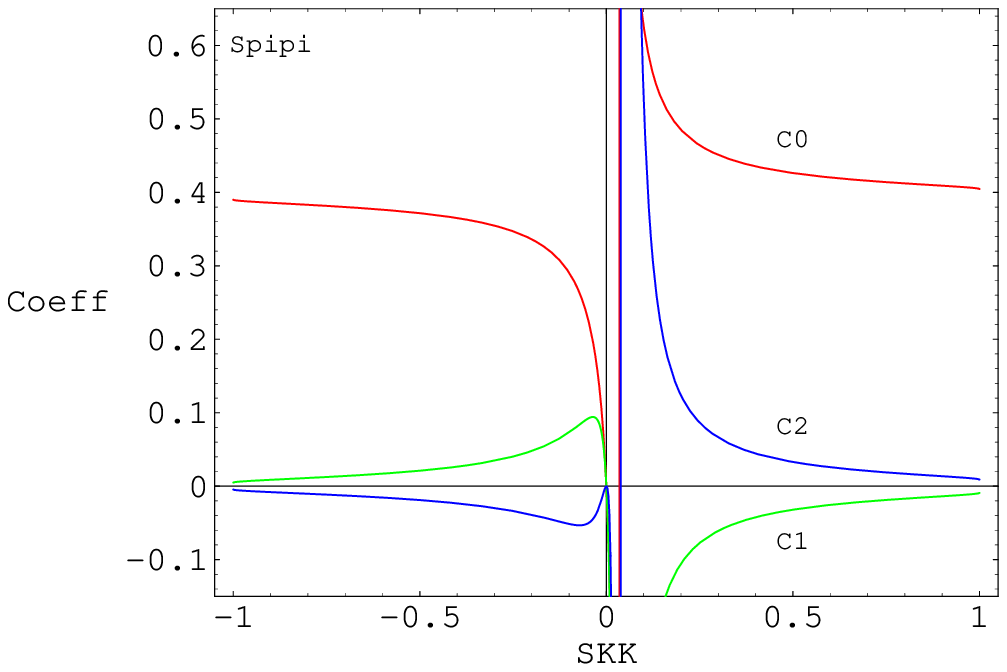,width=7.5cm,height=5cm}
\end{center}
{\it{\caption{The dependence of $\kappa_{0,1,2}$ on $S_{K K}$ for 
various values of $S_{\pi \pi}$.
\label{fig:etaSKK-coef}}}}
\end{figure}
In Fig.~\ref{fig:etaSKK}, we present the dependence of $\bar\eta$ as a function 
of $S_{K K}$ for various values of $S_{\pi \pi}$ .
The bands reflect the theoretical estimate on the  $SU(3)$-symmetry-breaking 
factor, $\zeta_{SU(3)}=1\pm0.3$. We observe that the sensitivity to the 
dependence on the $SU(3)$-symmetry-breaking factor is less pronounced, for 
$S_{K K}>0.4$. The reason for that could be easily understood  from 
Eq.(\ref{etabSU3}). To this end, it is more adequate to rewrite (\ref{etabSU3}), 
as a Taylor Expansion around its $SU(3)$-symmetry limit, namely 
$\zeta_{SU(3)}=1$, as :
\begin{eqnarray}\label{etabSU3-develop}
\bar\eta=\frac{1+\tau \Spipi-\sqrt{1 - \Spipi^2}}{(1 + \tau^2)\,\Spipi}
\Big[\sum_{i=0}^{2}\kappa_i~(\zeta_{SU(3)}-1)^i +{\cal O}(\zeta_{SU(3)}-1)^3\Big],
\end{eqnarray}
where the corresponding $\kappa_i$ coefficients are :
\begin{eqnarray}\label{etabSU3-develop-coef}
\kappa_0 &=& 1+\frac{\lambda^2 + \Delta}{1-\Delta},\\
\kappa_i &=& (-1)^{i}\frac{\lambda^2 + \Delta}{(1-\Delta)^{i+1}},~~~~~~~i\neq0,
\end{eqnarray}
with :
\begin{eqnarray}\label{etabSU3-develop-coef}
\Delta=\lambda^2 
\frac{\frac{- 2\,\tau + ( 1 + \tau^2)\,\SKK }{1-\tau \SKK - \sqrt{1 - \SKK^2}}}
{\frac{(1 + \tau^2)\,\Spipi}{1+\tau \Spipi-\sqrt{1 - \Spipi^2}}}
.
\end{eqnarray}

So far, we have just rewritten (\ref{etabSU3}) in a comprehensible way to 
understand the $SU(3)$-symmetry-breaking dependence of $\bar\eta$, as shown 
in Fig.~\ref{fig:etaSKK}. We notice that its dependence on the 
$SU(3)$-symmetry-breaking factor is rather mild in the region $\SKK > 0.4$. 
This 
behavior is not accidental, but reflects implicitly the $\kappa_i$ coefficients behaviors in terms of the mixing-induced $CP$ violation parameter $\SKK$. In fact, as shown in Fig.~\ref{fig:etaSKK-coef}, the $\kappa_i=0,1,2$ coefficients have the same order of magnitude when $\SKK < 0.2$, however this is not true anymore when $\SKK$ increases. In this case, we observe that the two coefficients $-\kappa_1$ and $\kappa_2$ decrease considerably between $0.2<\SKK<0.4$ until they tend to zero  with $\SKK > 0.4$. On the other hand, in the whole region $\SKK>0.2$, the coefficient $\kappa_0$ is the dominant one around a certain fixed value. Finally, we see that the insensitivity of $\bar\eta$ on the $SU(3)$-symmetry-breaking term in the region $\SKK>0.4$, is due mainly to the smallness of the two coefficients $\kappa_1$ and $\kappa_2$, which tend to vanish in this region.
\subsection{Estimate of $\zeta_{SU(3)}$ in QCD Factorization}
The analyses described above require mainly theoretical input on the $SU(3)$-symmetry-breaking effects $\zeta_{SU(3)}$, which is badly established at present~\cite{Khodjamirian:2003xk}. However, several approaches have been employed to predict the $SU(3)$-symmetry-breaking to the ratio of the kaon and pion decay constants $f_K/f_{\pi}$ and/or to the ratio of $B\to K$ and $B\to \pi$ form factors. Most of these theoretical approaches rely on 
QCD sum rules, Quark Model and Lattice QCD, leading to a quantitative 
estimate of $\sim 20\%$ in the $B\to P$ transition~\cite{Ball:1998tj}  
(which has been updated in \cite{Ball:2004ye}, leading to $\sim 30\%$)
and up to $\sim 30\%$ in the $B\to V$ sector~\citer{Ali:1995uy}, reflecting our present poor knowledge on its estimation. Therefore, we have relied on its generic value, namely $\sim 30\%$, to perform our analyses.
To reinforce the validity of this approximation, it is important to test  this 
estimation within the QCD factorization framework.
This was already discussed in \cite{Beneke:2003zw}.
Our aim is to reanalyze 
this quantity within the context of our phenomenological analysis.

In this approach, the $SU(3)$-symmetry-breaking effects do enter 
through different quantities, such as the ratio of the kaon and pion 
decay constants $f_K/f_{\pi}$, the ratio of 
$F_0^{B_s \to K}/F_0^{B \to \pi}$ form factors, the Gegenbauer moments 
$\alpha^K_i$ {\it vs.} $\alpha^{\pi}_i$, 
the first inverse moments of the $B$-meson distribution amplitudes 
$m_B/\lambda_B$ {\it vs.} $m_{B_s}/\lambda_{B_s}$ and phenomenological 
parameters $X_A$ and $X_H$ that enter power corrections to hard spectator scattering and weak annihilation effects.

As far as these parameters are poorly established, it would be interesting to 
exhibit their effect on the $SU(3)$-breaking term within our framework. In QCD factorization, the $SU(3)$-symmetry-breaking term is defined as:

\begin{eqnarray}\label{def-QCDFSU3}
\zeta^{\rm{QCDF}}_{SU(3)}&\equiv& \frac{r}{r'}
= \Bigg |
\frac{
\frac{a^c_4 + r^K_\chi a^c_6 + r_A[b_3+2 b_4]}{
 a_1+a^u_4 + r^K_\chi a^u_6 + r_A[b_1+b_3+2 b_4]}}
{\frac{a^c_4 + r^{\pi}_\chi a^c_6 + r'_A[b_3+2 b_4]}{
 a_1+a^u_4 + r^{\pi}_\chi a^u_6 + r'_A[b_1+b_3+2 b_4]}}
\Bigg |.
\end{eqnarray}
In estimating this quantity, we have used the input parameters in Table \ref{tab:input} with the additional one, needed for $B_d\to \pi^+ \pi^-$ modes, given in Table~\ref{tab:input-pipi}.

\renewcommand{\arraystretch}{1.3}
\begin{table}[t]
\begin{center}
\begin{tabular}{|c|c|c|c|c|}\hline\hline
$(m_u+m_d)~ (2~ {\rm GeV})$&
$f_B~({\rm MeV})$&
$F_0^{B\to\pi}(0)$&
$(\alpha_1^{\pi},~\alpha_2^{\pi})$&
$\lambda_B~({\rm MeV})$
\\ \hline
 $(9.1\pm 2.1)~({\rm MeV})$
& $180\pm 40$
& $0.28\pm 0.05$
& $(0,~0.1\pm 0.3)$
& $350\pm 150$
\\ \hline\hline
\end{tabular}
\end{center}
\centerline{\parbox{14cm}{\it{
\caption{\label{tab:input-pipi}
Additional input to Table \ref{tab:input} used in computing $\zeta^{QCDF}_{SU(3)}$.}}}}
\end{table}

In Table \ref{tab:zetaSU3} we show the value for $\zeta^{\rm{QCDF}}_{SU(3)}$
from a calculation within the QCD factorization framework
as described in \cite{BBNS3}. We also display the uncertainties
from various sources, distinguishing two classes.

The first corresponds to the impact of all the input parameters in 
Table~\ref{tab:zetaSU3}, apart from those in the last column, reflecting 
the second class of uncertainties. These latter category exhibits
the effects of hard spectator scattering and weak annihilation contributions, 
parametrized by  phenomenological quantities (see Eq.(\ref{xhxa})), on our 
$SU(3)$-breaking estimate within the QCD factorization framework. 
The default values have $\rho_{A(H)}=0$. Assuming the universality on the 
second class of uncertainties\footnote{namely $\rho_{A(H)}^{K}=\rho_{A(H)}^{\pi}=\rho_{A(H)}$ and $\phi_{A(H)}^{K}=\phi_{A(H)}^{\pi}=\phi_{A(H)}$.} and assigning an error of $100\%$ on $X_{A(H)}$ in (\ref{xhxa}), by allowing for arbitrary phases $\phi_{A(H)}$ and taking $\rho_{A(H)}$ between 0 and 1, implies the dominance of the annihilation contributions and the Gegenbauer moments of the kaon meson wave function among the remaining input parameters, as shown in Table~\ref{tab:zetaSU3}.
%
\renewcommand{\arraystretch}{}
\begin{table}[t]
\begin{center}
\begin{tabular}{|c|c|c|c|c|c|c|c|c|}\hline\hline
$\zeta
^{\rm{QCDF}}
_{SU(3)}$&
$\mu$&
$m_q$&
$m_c$&
$f_{B_{s[d]}}$&
$F_0^{s[d]}
$&
$\alpha_{1,2}^{K[\pi]}$&
$\lambda_{B}$&
$(\rho,\phi)_{H[A]}$
\\ \hline
  $1.03$
& $\pm 1.5$
& $\pm 3$
& $\pm 5$
& $\pm0.02$
& $\pm0.02$
& $\pm0.05$
& $\pm 1$
& $\pm0.001$
\\ 

& $\times 10^{-3}$
& $\times 10^{-3}$
& $\times 10^{-3}$
& $\pm0.03$
& $[\pm0.02]$
&$[\pm0.03]$
& $\times 10^{-3}$
&$[\pm0.06]$
\\ \hline\hline
\end{tabular}
\end{center}
\centerline{\parbox{14cm}{\it{
\caption{\label{tab:zetaSU3}
The QCD factorization predictions for $\zeta^{\rm{QCDF}}_{SU(3)}$ and its
uncertainties.
The labels $m_{q}$ and $F_0^{s[d]}$ denote respectively $m_s$, $m_u+m_d$ and
 $F_0^{B_{s[d]}\to K[\pi]}$.
}}}}
\end{table}
Adding the errors in quadrature, one obtains
\begin{eqnarray}\label{QCDFSU3}
\zeta^{\rm{QCDF}}_{SU(3)}=  1.03 \pm  0.07 \pm 0.06,
\end{eqnarray}
where the first (second) errors are from the first (second) class of 
uncertainties. Combining both in quadrature we finally get:
\begin{eqnarray}\label{QCDFSU3-f}
\zeta^{\rm{QCDF}}_{SU(3)}=  1.03 \pm  0.09.
\end{eqnarray}

In the case of non-universality of hard spectator scattering and weak annihilations terms\footnote{We mean here that $\rho_{A(H)}^{K}\neq\rho_{A(H)}^{\pi}$ and $\phi_{A(H)}^{K}\neq\phi_{A(H)}^{\pi}$.},
we pick the corresponding $X_{A,H}$ value for the pion mode and allow the corresponding parameters to vary by $\pm 30\%$ for the kaon mode. We find that the $SU(3)$-breaking effect on the ratio of penguin amplitudes (\ref{def-QCDFSU3}) can be large up to $ 30\%$ in magnitude, with a complete dominance from the annihilation contributions. We conclude that $\zeta^{\rm{QCDF}}_{SU(3)}$ is at present poorly determined by QCD Factorization. Therefore to get control over its estimate, a better understanding of the weak annihilation part is in order.

\section{Conclusions}
We have investigated the impact of the forthcoming measurements of the 
time-dependent $CP$ asymmetry parameters, namely $S_{KK}$ and $C_{KK}$ in 
$B_s\to K^+K^-$ decays, on the extraction of weak phases, which will soon 
become accessible at hadron machines, namely Tevatron Run-II, BTeV and LHC-b.
We shall conclude by summarizing our main results as follows.

An efficient use of mixing-induced $CP$ violation in $B_s\to K^+K^-$
decays, measured by $S_{KK}$, can be made by combining it with the 
corresponding observable from $B_d \to J/\psi K_S$, namely $\sin 2\beta$, 
the $B^0_s$--$\bar{B^0_s}$ mixing phase $\phi_s$ and the hadronic penguin 
parameters $(r, \theta)$ in extracting the CKM parameters 
$\bar\rho$ and $\bar\eta$. Looking more closely, to their dependences on
hadronic parameters, it turns out that the sensitivity
of $\bar\eta$ on the strong phase $\theta$ is very weak, which is entering
in $\bar\eta$ at second order, implying a negligible effect for its moderate 
values. Consequently a simple determination of the UT from $\tau$ and $S_{KK}$
is possible since the dependence of $\bar\eta$ on $r$ is proportional to  
an overall factor $(\rt-1)$.

Moreover, if the $B^0_d$--$\bar{B^0_d}$ mixing phase $\phi_d=2 \beta$ is fixed 
through $B_d \to J/\psi K_S$, then possible $CP$-violating NP contributions to 
$B_s^0$--$\bar B_s^0$ mixing may shift the range for the mixing-induced $CP$ 
violation observable $S_{KK}$ (but not $C_{KK}$)  significantly.
To illustrate this scenario, we have presented contour 
plots, allowing us to distinguish between the SM and NP scenario.
In this case, NP signal in the  $B^0_s$--$\bar{B^0_s}$ mixing would be 
provided by a sizeable contribution either from the angular distribution 
analysis of $B_s\to J/\psi \phi$ and $B_s\to D^{*+}_s D^{*-}_s$ decays or 
from the $CP$-violating asymmetries in $B_s\to J/\psi \eta^{(')}$ and 
$B_s\to D^{+}_s D^{-}_s$ systems.
%
In addition to $S_{KK}$, a complementary information from the direct $CP$ 
violation parameter $C_{KK}$ are investigated. Although its dependence
on hadronic input is more pronounced than for $S_{KK}$, it constraints 
significantly the dynamical quantities $r$ and $\theta$. Interestingly, we find
for the conservative bound $r \approx 0.15$ a strong constraint on $C \gtrsim- 0.26$ could 
be obtained, independently on the strong phase $\theta$.

As an input to the phenomenological discussion we also studied
the calculation of the penguin parameters $r$ and $\theta$ in QCD.
We have analyzed $r$ and $\theta$ within QCD factorization
with a particular view on theoretical uncertainties implying a large 
uncertainty due to the unknown annihilation contributions. An alternative way
 to explore these penguin parameters is using the plausible assumption based 
on the $SU(3)$-flavour-symmetry of strong interactions. Thus, one can complement the 
$CP$-violating observables of $B_s\to K^+K^-$ in a variety of ways with the 
current $B$-factories data provided by $B_d\to \pi^+\pi^-$ channels.
Interestingly, we show that the dependence of $r$ on $S_{K K}$ fixed through   the $CP$-violating observables $\Spipi$ and the $SU(3)$-symmetry-breaking 
factor $\zeta_{SU(3)}=1\pm0.3$,  
implies a positive value of $S_{K K}$ (see Fig.~\ref{fig:rSKK}).
Moreover, the sensitivity on the $SU(3)$-symmetry-breaking factor 
$\zeta_{SU(3)}$ in determining $\bar\eta$ as function of $S_{K K}$, 
fixed through  the $CP$-violating observable $S_{\pi \pi}$, is found 
rather mild for $\SKK>0.4$. To reinforce the validity of our approximation,
we have analyzed the $SU(3)$-symmetry-breaking factor $\zeta_{SU(3)}$ within QCD factorization with a particular view on theoretical uncertainties.



\subsection*{ Acknowledgements:} 
I am indebted to Gerhard Buchalla for helpful comments and discussion.
Fruitful conversations with Robert Fleischer are gratefully acknowledge.
This work is supported in part by the Deutsche 
For\-schungs\-ge\-mein\-schaft (DFG) under contract BU 1391/1-2. I would 
like to  thank the CERN Theory Group for the kind hospitality during my visit.


\vfill\eject

\end{document}